\documentclass[sn-mathphys-num,iicol,pdflatex]{sn-jnl}
\hypersetup{hidelinks}
\usepackage{xcolor}
\usepackage[normalem]{ulem}

\usepackage{graphicx}%
\usepackage{dcolumn}%
\usepackage{bm}%
\usepackage{physics}
\usepackage{bbm}
\usepackage{amsmath}
\usepackage{amsfonts}
\usepackage[capitalise]{cleveref}

\newcommand{\vr}[0]{{{\bf r}}}
\newcommand{\vp}[0]{{{p}}}
\newcommand{\mA}[0]{{{\bm A}}}

\newcommand{\phat}[0]{\hat{p}}
\newcommand{\bmphat}[0]{\hat{\vp}}

\newcommand{\Vall}[0]{V_{\textnormal{all}}(\vr_1,\vr_2,...,\vr_N)}

\begin{document}
% \title{Local Pseudopotentials Empower Neural Network-based Quantum Monte Carlo}
\title{Empowering Neural Network-based Quantum Monte Carlo with Local Pseudopotentials}

\author[1, 2]{\fnm{Weizhong} \sur{Fu}}
\equalcont{These authors contributed equally to this work.}

\author[3]{\fnm{Ryunosuke} \sur{Fujimaru}}
\equalcont{These authors contributed equally to this work.}

\author[1]{\fnm{Ruichen} \sur{Li}}
\equalcont{These authors contributed equally to this work.}

\author[1]{\fnm{Yuzhi} \sur{Liu}}
\equalcont{These authors contributed equally to this work.}

\author[1]{\fnm{Xuelan} \sur{Wen}}

\author[1]{\fnm{Xiang} \sur{Li}}

\author[4]{\fnm{Kenta} \sur{Hongo}}

\author*[5, 6]{\fnm{Liwei} \sur{Wang}}
\email{wanglw@pku.edu.cn}

\author*[3]{\fnm{Tom} \sur{Ichibha}}
\email{ichibha@gmail.com}

\author*[7]{\fnm{Ryo} \sur{Maezono}}
\email{rmaezono@mac.com}

\author*[2, 8, 9]{\fnm{Ji} \sur{Chen}}
\email{ji.chen@pku.edu.cn}

\author*[1]{\fnm{Weiluo} \sur{Ren}}
\email{renweiluo@bytedance.com}

\affil[1]{ByteDance Seed}
\affil[2]{School of Physics, Peking University, Beijing 100871, People's Republic of China}
\affil[3]{School of Information Science, JAIST, Asahidai 1-1, Nomi, Ishikawa 923-1292, Japan}
\affil[4]{Center for Advanced Scientific Computing, JAIST, Asahidai 1-1, Nomi, Ishikawa 923-1292, Japan}
\affil[5]{National Key Laboratory of General Artificial Intelligence, School of Intelligence Science and Technology, Peking University, Beijing, P. R. China}
\affil[6]{Center for Machine Learning Research, Peking University, Beijing, P. R. China}
\affil[7]{Graduate Major in Materials and Information Sciences, School of Materials and Chemical Technology, Institute of Science Tokyo, 2-12-1-S6-22 Ookayama, Meguro-ku, Tokyo 152-8550, Japan}
\affil[8]{Interdisciplinary Institute of Light-Element Quantum Materials and Research Center for Light-Element Advanced Materials, Peking University, Beijing 100871, People's Republic of China}
\affil[9]{State Key Laboratory of Artificial Microstructure and Mesoscopic Physics and Frontiers Science Center for Nano-Optoelectronics, Peking University, Beijing 100871, People's Republic of China}

\abstract{
Neural Network-based Quantum Monte Carlo (NNQMC), an emerging method for solving many-body quantum systems with high accuracy, has been mainly applied to small systems due to demanding computation requirements. In this work, we introduce a framework based on local pseudopotentials to break through such limitation, improving the computational efficiency and scalability of NNQMC. The incorporation of local pseudopotentials reduces the number of electrons treated in neural network and also achieves better relative energy accuracy than all electron NNQMC calculations for complex systems. This counterintuitive outcome is made possible by the distinctive characteristics inherent to NNQMC. Notably, by avoiding costly integration terms, this approach is also substantially more efficient than its widely used semilocal counterparts. Our approach enables the reliable treatment of large and challenging systems, such as the $\text{Fe}_4 \text{S}_4 (\text{SCH}_3)_4$ iron-sulfur cluster. Overall, our findings demonstrate that the synergy between NNQMC and local pseudopotentials substantially expands the scope of accurate ab initio calculations.
}

\maketitle

\section{Introduction}
Solving Schr\"{o}dinger equation exactly is intractable for systems with more than a few particles due to the exponential growth of the Hilbert space with system size. 
As a practical alternative, Quantum Monte Carlo (QMC)~\cite{needs2020variational, foulkes2001quantum} methods offer a well-established approximate approach for solving the Schr\"{o}dinger equation with high accuracy and relatively low computational cost.
Building on the QMC framework, Neural Network-based Quantum Monte Carlo (NNQMC)~\cite{han2019solving, pfau2020ab, spencer2020better, hermann2020deep, vonself,ren2023towards, li2022ab, li2024computational} has recently emerged as a transformative extension that harnesses the power of neural networks to represent wavefunctions.
By leveraging the expressive power of deep neural networks to represent wavefunctions, NNQMC offers a highly accurate approach to solving many-body quantum problems.
However, its achievable system size remains limited by high computational costs from optimizing numerous parameters and prolonged training for convergence.

Various improvements have been proposed to address this issue.
For instance, joint training frameworks, together with transferrable wavefunction models across different systems, eliminate the need for individual training of each configuration~\cite{scherbela2022solving, scherbela2023towards, scherbela2023variational, gaoab, gaosampling, gao2023generalizing, gerard2025transferable}.
On the other hand, a Forward Laplacian approach combined with efficient neural network design significantly accelerates the calculation of laplacian operator, which is the computational bottleneck in NNQMC~\cite{li2024computational}.
Also, a recently proposed finite-range embedding approach achieves substantial efficiency improvement by considering only nearby electron interactions~\cite{FiRE}.
However, applying NNQMC to large systems in practical studies remains challenging.

In this work, we scale NNQMC to much larger, strongly correlated systems through integration with local pseudopotentials~\cite{bachelet1989novel, foulkes1990pseudopotentials}.
This approach achieves significantly faster NNQMC workflows than all-electron or semilocal-pseudopotential calculations without any accuracy compromise for relevant chemical observables.
High-quality local pseudopotentials are already available for a number of transition metals~\cite{bennett2022high, ichibha2023locality}; here we expand this coverage to three period-3 elements, phosphorus, sulfur and chlorine, demonstrating their effectiveness for light main group elements.
Additionally, this newly constructed local pseudopotential also enables us to study large iron-sulfur clusters, simulating different valence states and verifying the magnetic coupling constants.
Moreover, this technique can be seamlessly integrated with previous methological advances to further extend the applicability of NNQMC.
Taken together, these findings show that integrating local pseudopotentials can unlock the potential of NNQMC, especially as they extend to more elements in the future.
This is also much like pseudopotentials' critical role in 
helping density functional theory and first principles simulation take flight~\cite{bachelet1982pseudopotentials, vanderbilt1990soft}.

\section{Results}
\subsection{Framework Overview and Comparative Analysis}
\begin{figure*}[t]
    \centering
    \includegraphics[width=\linewidth]{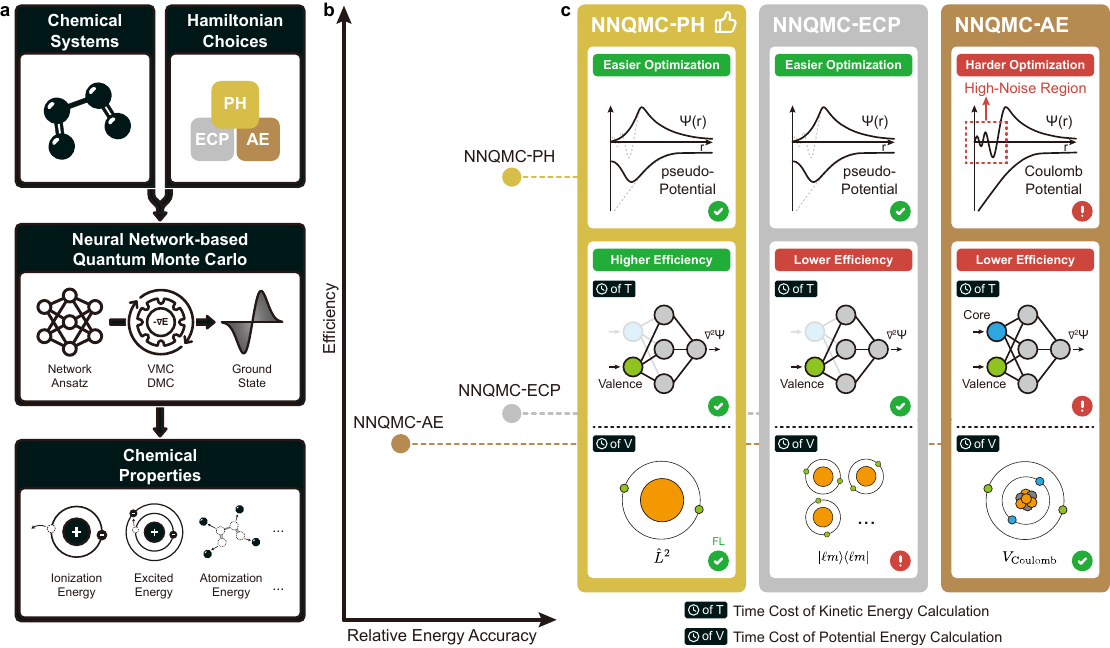}
    \caption{\textbf{Advantages of pseudo-Hamiltonians in neural network-based quantum Monte Carlo.} \textbf{a}~The overall framework. \textit{Top}: The studied chemical systems and three Hamiltonian form choices, that is, PH, ECP and AE, among which PH achieves the best performance. \textit{Middle}: The workflow of NNQMC. The wavefunction is represented by a neural network ansatz, optimized (VMC) or projected (DMC) towards minimal energy, and converged to the ground state. \textit{Bottom}: Different types of relative energies and chemical properties, including ionization energy, excited energy, atomization energy and more. 
    \textbf{b}~Efficiency and accuracy comparison of NNQMC using PH, ECP and AE Hamiltonians (for demonstration purposes only and is not to scale). Adopting PH achieves superior accuracy on calculating relative energies at significantly reduced computational cost. %
    \textbf{c}~Sources of PH's dual advantages. \textit{Top}: Pseudopotentials eliminate the wavefunction fluctuations in core-region caused by nuclear Coulomb singularities in AE, thus making wavefunctions under PH or ECP Hamiltonians easier to be optimized. \textit{Bottom}: The computational bottleneck lies in the energy calculation, consisting of kinetic and potential terms. By eliminating core electrons, PH/ECP substantially accelerates kinetic energy calculations. For potential energy, PH involves computing the angular momentum squared operator ($\hat{L}^2$), requiring second-order derivatives, whereas ECP includes nonlocal terms $\ket{\ell m}\bra{\ell m}$ based on spherical harmonics, necessitating costly integrations. Kinetic energy calculation acceleration combined with avoidance of ECP's nonlocal spherical integrals makes PH uniquely efficient. Additionally, with Forward Laplacian (FL) framework~\cite{li2024computational, li2024dof}, $\hat{L}^2$ in PH potential terms incurs almost no extra computational cost.}
    \label{fig:framework}
\end{figure*}

NNQMC framework enables accurate calculation of molecular ground state energies and chemical properties by integrating the neural network wavefunction ansatz (for instance, LapNet~\cite{li2024computational}) with QMC approaches, 
as illustrated in \cref{fig:framework}a.
To obtain the ground state,  variational Monte Carlo (VMC) minimizes the energy via gradient-based optimization.
For further refinement, diffusion Monte Carlo (DMC) projects the wavefunction toward the ground state by stochastically sampling imaginary-time evolution.
This workflow enables accurate predictions of various chemical properties sensitive to electron correlation effects, including ionization energy, excited energy, atomization energy and so on.
See \cref{NNQMC} for more theoretical details.

A key strength of NNQMC lies in its ability to accommodate a wide variety of Hamiltonians.
Specifically, it allows direct incorporation of different pseudopotentials, which are widely used in quantum chemistry, within a unified framework.
We consider different forms of the Hamiltonian in this work, including the all-electron (AE) ab initio Hamiltonian and one employing semi-local pseudopotentials, also known as effective core potentials (ECPs)~\cite{dolg1987energy,burkatzki2007energy, 2008MB_MD, trail2017shape,annaberdiyev2018new}. 
In this work, we propose incorporating local pseudopotentials, referred to as pseudo-Hamiltonians (PHs)~\cite{bachelet1989novel, foulkes1990pseudopotentials, ichibha2023locality, bennett2022high}, within the NNQMC framework. 
Those Hamiltonian choices are depicted at the top of \cref{fig:framework}a.

We briefly describe the differences between these Hamiltonian here (see illustration in \cref{fig:framework}c), with a detailed discussion provided in \cref{three_H}.
AE explicitly treats all electrons and nuclei via the Coulomb potential.
ECP replaces core electrons with smoothed potentials but introduce nonlocal angular momentum projectors $\ket{\ell m}\bra{\ell m}$.
PH, in contrast, combines pseudopotentials with a simplified angular momentum operator squared $\hat{L}^2$ in its potential terms, eliminating the need for computationally demanding spherical harmonics integrations.
Adopting effective Hamiltonians like ECP and PH maintains the broad applicability of NNQMC for chemical applications.
This is because the properties of interest are governed primarily by valence electrons, while the replaced core electrons are chemically inert~\cite{pauling1960nature,hay1985ab,kahn1976b}.
Furthermore, the relative energies calculated from those effective Hamiltonian are directly comparable to AE calculations and experimental measurements.
As we sketch in \cref{fig:framework}b, PH should be the preferred choice in NNQMC framework, which will be demonstrated with concrete benchmarks in the following subsections.

In this work, we assess accuracy using relative energies, comparing against experiments and other computational benchmarks.
This choice reflects the fact that chemically relevant quantities, such as ionization energy and atomization energy, are mostly relative energies, calculated as energy differences, rather than absolute energies~\cite{szabo2012modern,pople1989gaussian}.
For instance, the ionization energy is the energy difference between a neutral molecule and its corresponding ion, and atomization energy is the one between a molecule and its isolated constituent atoms~\cite{seth2011quantum,harkless2000quantum}.

Note that, the formally exact AE Hamiltonian does not always deliver the most accurate NNQMC results, especially when involving heavier elements~\cite{li2022fermionic}.
All-electron QMC suffer from the more challenging numerical behaviour, where rapidly varying and singular core-region features can destabilize optimization and hinder convergence~\cite{ceperley1995quantum,hammond1987valence}.
By replacing the true ionic potential with a softened version, pseudopotentials smooth these singularities, as illustrated in the upper part of \cref{fig:framework}c. 
This mitigates numerical instabilities and creates a better-behaved landscape for wavefunction optimization, enabling more reliable energy convergence, as further discussed in \cref{fig:eff}c.
Moreover, the all-electron ground-state wavefunction is higher-dimensional and contains high-frequency components near the nuclei, which are notoriously more difficult for neural networks to learn~\cite{xu2019frequency,zhang2025shallow}. 
Owing to these combined difficulties, all-electron NNQMC calculations, despite using the formally exact Hamiltonian, can yield much worse practical accuracy than PH and ECP when calculating for relative energies, as we will demonstrate in \cref{sec:acc}.

As for efficiency, the computational bottleneck in NNQMC is energy calculation, comprising kinetic and potential energy terms.
For kinetic energy, by removing core electrons, both PH and ECP drastically reduces the computation cost.
For potential energy, ECP requires nonlocal spherical harmonics projections $\ket{\ell m}\bra{\ell m}$, which is computationally expensive especially when neural networks are used to model the wavefunction~\cite{li2022fermionic}.
In contrast, PH replaces these costly nonlocal terms with the angular momentum squared operator ($\hat{L}^2$), where second-order derivatives, the most time-consuming component, incur negligible additional cost under the Forward Laplacian framework (see details in \cref{three_H,sec:fl} and Supplementary Note 2).
Due to these two reasons, PH uniquely achieves superior efficiency, as illustrated in the lower part of \cref{fig:framework}c.

Moreover, PH implementation is much simpler than ECP, especially when integrated with DMC (as ECP requires approximations like T-move~\cite{casula2006beyond, anderson2021nonlocal} or DLA~\cite{hammond1987valence, zen2019new}), and can be seamlessly integrated to different types of neural network ansatz~\cite{han2019solving,pfau2020ab,hermann2020deep, spencer2020better, FiRE, li2024computational, vonself, gaoneural, gerard2022gold} and variants of NNQMC such as WQMC~\cite{neklyudov2023wasserstein}.
The details of PH implementation within the Forward Laplacian framework are provided at \cref{sec:fl}.

\subsection{Major Efficiency Improvement}\label{sec:eff}
\begin{figure*}[t]
    \centering
    \includegraphics[width=\linewidth]{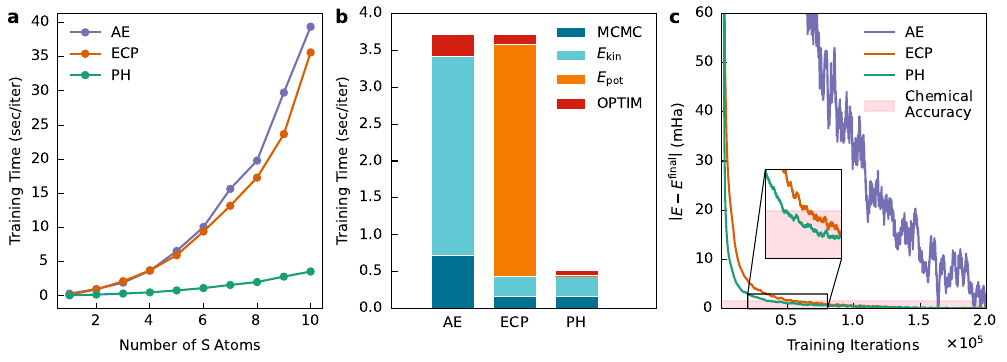}
    \caption{\textbf{NNQMC efficiency with AE, ECP and PH Hamiltonians.} \textbf{a}~Training time cost of NNQMC with the three Hamiltonians in pure sulfur systems with respect to the number of S atoms on 8 V100 cards. \textbf{b}~Detailed training time cost on S$_4$ system of different parts, that are respectively electron configurations sampling by Markov Chain Monte Carlo (MCMC), calculation of kinetic energy $E_{\rm kin}$ and potential energy $E_{\rm pot}$, and wavefunction optimization by KFAC optimizer~\cite{martens2015optimizing,kfac-jax2022github}. \textbf{c}~The three methods' convergence speed of energy calculation on S$_4$ system. $E^{\mathrm{final}}$ represents the energy corresponding to the final training iteration ($2\times10^5$ iterations), and the three energy curves are smoothed with a 2000-step rolling window. The inset zooms in the PH and ECP curves.}
    \label{fig:eff}
\end{figure*}

Initially designed to address locality errors in traditional QMC~\cite{ichibha2023locality, bennett2022high}, PH proves even more transformative for efficiency improvement when integrated with NNQMC.
Specifically, PH reduces the number of electrons involved in calculations with considerably less overhead compared to its semi-local counterpart, ECP.
In addition to transition metals~\cite{ichibha2023locality,bennett2022high}, period-3 elements are particularly well-suited for this technique due to their high proportion of core electrons relative to total electrons, with sulfur being a notable example.
In this work, we construct PH for three period-3 elements, hosphorus, sulfur and chlorine (See \cref{PH:construct} for details), achieving state-of-the-art efficiency and accuracy.
To put this into perspective, now we are able to simulate iron-sulfur cluster ($\text{Fe}_2\text{S}_2\text{(SH)}_4$) with the same computational resources for benzene dimer when using PH for both iron and sulfur atoms.

We demonstrate the efficiency improvement using a series of sulfur-only systems in \cref{fig:eff}a, ranging from a single sulfur atom to S$_{10}$.
Across all these systems, ECP calculations incur computational cost comparable to AE, while PH is able to achieve more than 10 times acceleration for larger systems.
Furthermore, we analyze contribution of each component in calculation runtime, using S$_4$ as an example.
As shown in \cref{fig:eff}b, AE and ECP runtimes are both significantly greater than PH, but for different reasons: AE is dominated by kinetic energy due to the large number of involved electrons, while ECP is dominated by potential energy due to the nonlocal terms.
In contrast, PH has a negligible overhead in terms of potential energy, and runtimes for both kinetic energy and MCMC are also improved due to the reduced number of involved electrons.
Additional systematic analyses across a broader set of systems are provided in Supplementary Note 3.

In addition to per-iteration efficiency improvement, the overall convergence of the whole training process is also improved with PH, especially compared to AE, as shown in \cref{fig:eff}c.
In particular, both ECP and PH converges after around 50,000 training iterations, namely within the chemical accuracy of the last-step  value, while AE is far from convergence even after 150,000 iterations.
Notably, the training curve for AE is also fluctuating more severely, which is expected due to the inclusion of the core electrons.

In summary, PH generally enhances NNQMC's computational efficiency by over tenfold, enabling complex systems simulations with reduced resources. 
More importantly, this efficiency boost comes along with notable improvements in accuracy, as discussed in the following subsection.

\subsection{Accuracy Benchmarks}\label{sec:acc}
\begin{figure*}[t]
    \centering
    \includegraphics[width=140mm]{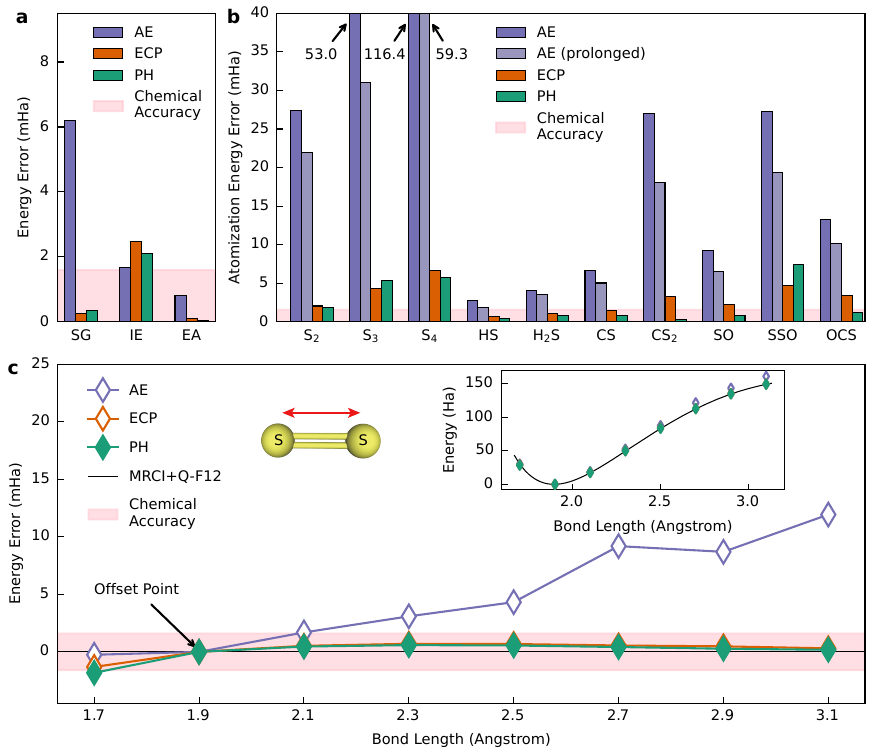}
    \caption{\textbf{NNQMC accuracy with AE, ECP and PH Hamiltonians.} \textbf{a}~The energy error of spin gap (SG), ionization energy (IE) and electron affinity (EA) of S atom. The benchmarks are experimental results~\cite{martin1990energy, kelly1987atomic, blondel2005electron}. \textbf{b}~The atomization energy error of various sulfur-containing molecules. In addition to results obtained with the default training length ($2\times10^5$ iterations), we also present AE results from prolonged training ($1\times10^6$ iterations) as a better-converged reference. The benchmarks are results of W4 database~\cite{karton2011w4}. \textbf{c}~The potential energy curve errors of the sulfur dimer with different Hamiltonians. The energies at 1.9\AA~serve as reference points for energy offsets. The benchmark is the results of MRCI+Q-F12 under aug-cc-pV(5+d)Z basis set~\cite{sarka2023potential}. Inset: The potential energy curves. In all the calculations with pseudopotentials (ECP or PH), they are applied only on sulfur atoms. %
    }
    \label{fig:acc}
\end{figure*}

Pseudopotentials improve NNQMC accuracy by enabling a more stable optimization process and better overall convergence, achieved through the removal of troublesome core electrons.
Specifically, PH calculations yield results consistent with ECP, significantly outperforming AE in problems involving heavier elements.
It is worth noting that because AE, ECP and PH employ different Hamiltonians, their absolute energies cannot be directly compared.
Instead, the accuracy discussed here refers to the agreement of relative energies, such as atomization energies, with experimental results or other benchmarks.

As a first validation, we benchmark PH on spin gap (SG), ionization energy (IE), and electron affinity (EA) of the sulfur atom against experimental references~\cite{martin1990energy, kelly1987atomic, blondel2005electron}.
As shown in \cref{fig:acc}a, PH shows excellent agreement with ECP and achieves near-exact accuracy for the spin gap and EA relative to experiment, while AE performs poorly on the spin gap.
The only noticeable error of PH and ECP appears in the IE, likely due to the 1.8 mHa discrepancy between the CCSD(T) result (378.9 mHa) and the experimental result (380.7 mHa), as the CCSD(T) result served as the reference value during ECP/PH construction.
To further enhance accuracy, refinements to the ECP/PH construction procedure are needed, as discussed in \cref{sec:discuss}.
Moreover, we compute atomization energies for a variety of sulfur-containing molecules. 
As demonstrated in \cref{fig:acc}b, PH achieves chemical accuracy for most species and shows strong consistency with ECP, often delivering slightly better results.
In contrast, AE underperforms for most species, particularly in large molecules such as S$_3$ and S$_4$.
Since AE calculations do not converge well within the default training length (see \cref{fig:eff}c as an example for the S$_4$ molecule, and all full training curves in Supplementary Note 4), we also present AE results obtained with a fivefold longer training (light purple bars).
Although this extended training improves accuracy across all systems, the overall conclusion remains unchanged.
Even in the smallest system (HS) the AE accuracy (despite consuming roughly 14 times more computation resources) still underperforms compared to both ECP and PH.
Furthermore, we generate the potential energy curve (PEC) of the sulfur dimer, as shown in \cref{fig:acc}c.
To assess the non-parallelity errors (NPEs), we reference PECs to energies at the bond length of 1.9 \AA, with benchmark data from MRCI+Q-F12/aug-cc-pV(5+d)Z calculations~\cite{sarka2023potential}.
The PH and ECP PECs align closely, both exhibiting low NPEs (respectively 2.4 and 2.0 mHa).
AE, however, shows a large NPE (12.2 mHa), mostly rising from the energy results at bond lengths of 2.7-3.1 \AA, a crossover region between triplet and singlet states, where strong correlation effects make accurate modeling difficult (see Supplementary Note 5 for more details).

\subsection{Framework Generality}
\begin{figure}[t]
\centering
\includegraphics[width=\linewidth]{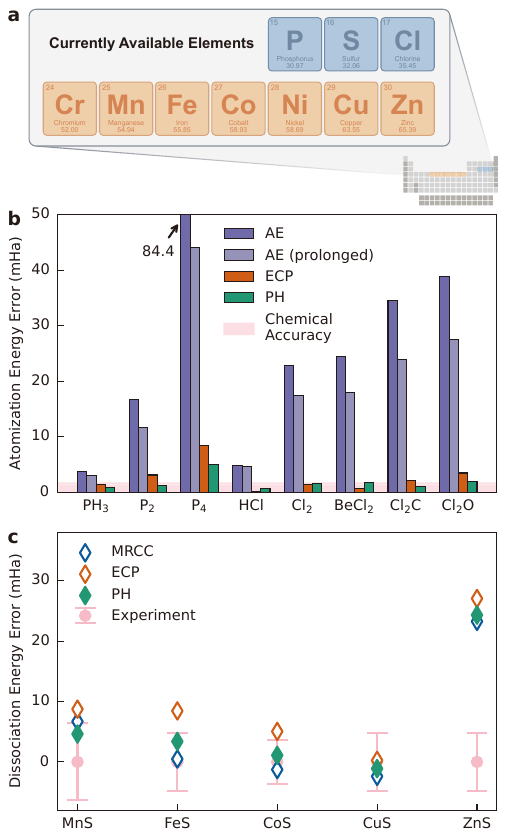}
    \caption{\textbf{Generalizability of NNQMC-PH to phosphorus-containing clusters, chlorine-containing clusters and transition-metal sulfides.} \textbf{a} Current element coverage of high quality PHs. \textbf{b}~The atomization energy error of various molecules containing phosphorus or chlorine. The AE and prolonged AE results correspond to  default training length ($2\times10^5$ iterations) and $1\times10^6$ iterations respectively. The benchmarks are results of W4 database~\cite{karton2011w4}. \textbf{c}~The dissociation energy error of various transition metal sulfides, with the MRCC results for comparison and experimental results as benchmark~\cite{aoto2017arrive}. In all the calculations with pseudopotentials (ECP or PH), they are applied only on phosphorus, sulfur, chlorine and transition metal atoms. }
    \label{fig:generality}
\end{figure}

The formulation of NNQMC with PH framework is general. 
As a general pseudopotential form, PH is independent of the downstream electronic-structure solver, and can be constructed for a broad range of elements given suitable reference data.
Within NNQMC, PH enters solely through the Hamiltonian evaluation and thus integrates without disrupting the overall workflow.
This makes PH compatible with other NNQMC advances, such as improved ansatz designs, accelerated differential operators, and transferable training paradigms. 
Together these advances can push the boundaries of NNQMC to previously intractable challenges.

In practice, the performance of our framework depends on the quality of the fitted PH, as in any pseudopotential-based approach.
In this regard, transferability is essential, which requires a PH constructed for a given element to provide a reliable valence-only effective Hamiltonian across diverse molecules and chemical environments. 
Achieving high transferability guides our fitting procedure, and in \cref{sec:acc} we demonstrated strong transferability of the resulting sulfur PH.
Here we expand the PH coverage to two additional period-3 elements, phosphorus and chlorine, using the same fitting protocol, which further supports generality for high quality PH. 
Together with the high quality PHs constructed previously for transition metals in Ref.~\cite{bennett2022high,ichibha2023locality}, we now have high quality PHs for 10 elements in total, as shown in \cref{fig:generality}a. 
Benchmarks on atomization energies for representative molecules containing phosphorus or chlorine (\cref{fig:generality}b) show competitive performance, with most cases achieving chemical accuracy.
The full training curves are displayed in Supplementary Note 4 as well.
Because PH removes the same number of core electrons from NNQMC calculation for period-3 elements, the efficiency gains for phosphorus and chlorine are similar to those for sulfur, as detailed in Supplementary Note 3.

As a more challenging task, we apply NNQMC with ECP/PH to transition metal sulfides (TMSs), which are far beyond AE's capabilities due to the heavy metal atoms. 
Specifically, we carry out calculation on five different dimers with reliable experimental benchmarks~\cite{aoto2017arrive} and the results are shown in \cref{fig:generality}c.
For PH calculations, all elements (both transition metals and sulfur) are treated with PH, and similarly for ECP calculations.
Our PH results are close to ECP as well as multi-reference coupled cluster (MRCC), another accurate quantum chemistry method.
Moreover, our PH results are within the experimental uncertainties, except for ZnS, probably due to the same reason as the deviated result of the sulfur atom's IE discussed earlier.
Given PH's success with TMS systems, we propose applying it to study some more meaningful systems, such as iron-sulfur clusters, as elaborated in the next subsection.

\subsection{Iron-Sulfur Clusters}
\begin{figure*}[t]
\centering
\includegraphics[width=1.0\linewidth]{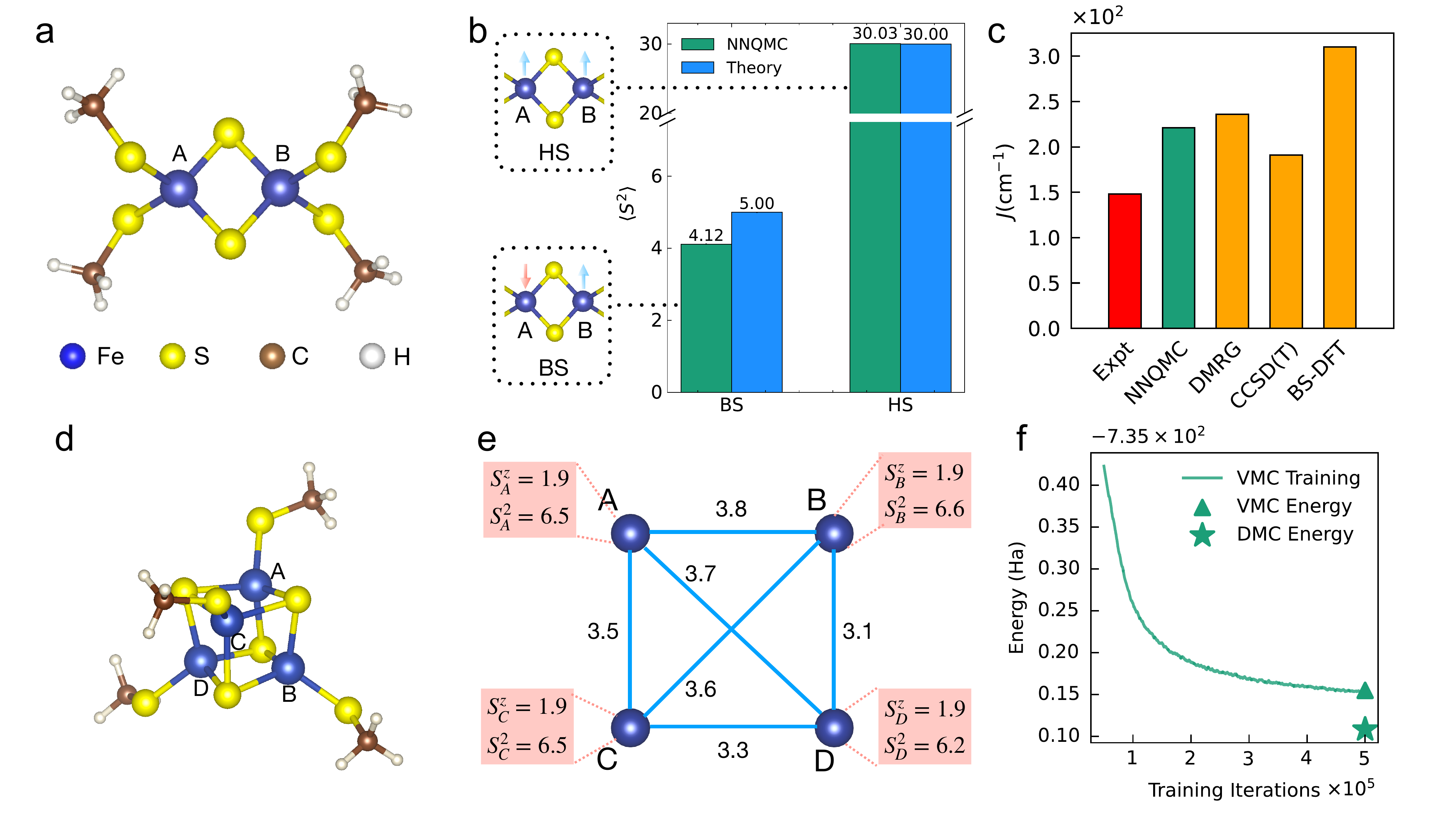}
    \caption{\textbf{Iron-sulfur cluster study using NNQMC with PH.} \textbf{a}~molecular structure of $[\text{Fe}_2 \text{S}_2 (\text{SCH}_3)_4]^{2-}$.
    The balls of different colors represent different elements: blue is iron, yellow is sulfur, brown is carbon, and white is Hydrogen. 
    The molecular structure is the same as \citet{sharma2014low}. 
    \textbf{b}~Theoretical and calculated values of spin squared ($S^{2}$) for the LS and HS states, along with the spin distribution of the LS and HS states on each iron (Fe) atom.
    \textbf{c}~Comparison of magnetic exchange coupling constants $J$ obtained from different approaches~\cite{schurkus2020theoretical, sharma2014low, noodleman1992density}. 
    The red bar is experimentally fitted value~\cite{haselhorst1993mu} and the green bar is from our NNQMC with PH calculations.
    \textbf{d}~Molecular structure of the cubane-type cluster $[\text{Fe}_4 \text{S}_4 (\text{SCH}_3)_4]$~\cite{li2021resolution} with the color same as \textbf{a}.
    \textbf{e}~Spin related information for the simulated HS state of the cubane-type cluster using NNQMC with PH. 
    The values on connect lines represent the spin correlation $\left<S_{i}S_{j}\right>$ between the iron sites being connected.
    Local $S^{z}_{i}$ and $S^{2}_{i}$ are presented beside each iron site. 
    \textbf{f}~The training curve for the HS state of the cubane-type cluster using NNQMC with PH, together with the VMC inference value and the DMC result for the last training iteration. }
    \label{fig:fes}
\end{figure*}
    
Iron-sulfur clusters, ubiquitous in biological systems and industrial catalysis, play crucial roles in enzymatic reactions and energy conversion processes.
Their intricate spin architectures, governed by superexchange interactions, have motivated extensive research efforts in quantum chemistry~\cite{beinert1997iron, noodleman1995orbital, noodleman1981valence}. 
With PH, we are now able to use NNQMC to analyze the spin configurations of two prototypical complexes: dinuclear $[\text{Fe}_2 \text{S}_2 (\text{SCH}_3)_4]^{2-}$ and cubane-type $[\text{Fe}_4 \text{S}_4 (\text{SCH}_3)_4]$, whose structures are depicted in \cref{fig:fes}a and d, respectively.
In these systems, PH achieves over 20 times acceleration.

For dinuclear iron-sulfur cluster $[\text{Fe}_2 \text{S}_2 (\text{SCH}_3)_4]^{2-}$, we focus on its magnetic exchange coupling constant $J$, which is a critical descriptor of magnetic interaction and thus an important parameter in quantum Heisenberg model~\cite{arovas1988functional}.
We calculate this constant following a broken-symmetry approach~\cite{schurkus2020theoretical, noodleman1981valence},
\begin{equation}
J = \frac{E_{\mathrm{BS}} - E_{\mathrm{HS}}}{\langle S^2\rangle _{\mathrm{BS}} - \langle S^2\rangle_{\mathrm{HS}}},
\label{eq:BS}
\end{equation}   
where BS denotes the broken-symmetry state and HS denotes the high-spin state.
The BS state is obtained by specifying a broken-symmetry unrestricted Hartree-Fock state as the pretraining target~\cite{lykos1963discussion}.
As shown in \cref{fig:fes}b, the $\langle S^2\rangle$ results for the two states are respectively 4.12 and 30.03, consistent with theoretical predictions and confirming their identification as BS and HS states.
The derived constant $J$ ($221\text{cm}^{-1}$)  agrees well with both experimental measurements and high-level theoretical methods, including DMRG and CCSD(T)~\cite{sharma2014low, schurkus2020theoretical}, as shown in \cref{fig:fes}c. 
This underscores the capacity of NNQMC with PH to deliver quantitative insights into magnetic interactions in strongly correlated systems.

To further demonstrate NNQMC with PH on more challenging systems, we simulate the HS states of cubane-type cluster $[\text{Fe}_4 \text{S}_4 (\text{SCH}_3)_4]$. 
This cluster has 268 total electrons and 148 valence electrons, which was previously beyond the reach of NNQMC methods.
The spin structure from our PH calculation is shown in \cref{fig:fes}e.
The spin correlation between the iron sites $\left<S_i S_j\right>$ ranges from $3.1$ to $3.8$, indicating ferromagnetic.
On each iron site, the local spin squared $S^{2}_{i}\approx6.5$ and the local spin along z-axis $S^{z}_{i}\approx1.9$, suggesting uniform spin distribution across the Fe atoms.
Empirically, a simple model to describe the HS state of $[\text{Fe}_4 \text{S}_4 (\text{SCH}_3)_4]$ is the Heisenberg model. 
Here we find that it indeed predict qualitatively correct results, but quantivative discrepancies arise from the model's neglect of charge/spin transfer effects (see further details in Supplementary Note 9).  
Our ab initio calculations can better capture the magnetic behavior, as similarly discussed in \citet{sharma2014low}.

This cubane-type cluster also serves as a good showcase of the integration between neural network-based DMC and PH, which offers advantages such as straightforward implementation and being free of locality errors \cite{bennett2022high, ichibha2023locality}.
We employ fixed-node DMC following 500,000 iterations of VMC training process, a reasonable number given this system’s large size and strong correlation.
As shown in \cref{fig:fes}f, the VMC training process is stable with small fluctuations, after which we obtain a HS state with a physically meaningful spin distribution on the iron sites, as previously discussed.
In light of prior studies \cite{ren2023towards}, we expect that the neural network wavefunction here provides an accurate nodal surface that enables DMC to approach the ground state of this challenging system, with a variational energy approximately 50 mHa lower than VMC.

\section{Discussions}
\label{sec:discuss}
This work presents a significant step forward in scalable, accurate quantum simulations by introducing the NNQMC with PH framework. 
The fully local nature of PH, with the help of the Forward Laplacian framework, eliminates the computational bottlenecks introduced by semi-local pseudopotentials.
As a result, this approach achieves state-of-the-art accuracy while dramatically improving efficiency, successfully extending simulations to systems such as the $\text{Fe}_4 \text{S}_4 (\text{SCH}_3)_4$ iron–sulfur cluster, which were previously intractable for NNQMC.

While the NNQMC with PH framework is in principle general, its practical applicability depends on the availability of high-quality PHs for specific elements.
Transition metals and period-3 elements, as demonstrated here, benefit substantially from PHs in both computational efficiency and optimization stability.
The advantages extend to lighter elements as well, especially in terms of optimization behavior, albeit more modestly because fewer core electrons are removed.
In contrast, ECPs face more serious issues in this regime and can even reduce efficiency rather than improve it.
On the other hand, applying NNQMC with PH framework to heavier elements like lanthanides and actinides will require additional developments in real-space QMC to better handle stronger relativistic effects and spin-orbital coupling. 
Although the element coverage of PH is still limited so far compared to mainstream ECPs~\cite{trail2017shape,annaberdiyev2018new}, we expect rapid improvement in the near future, motivated by our demonstration of the strong synergy between NNQMC and PH.
In collaboration with the community, we will further develop the NNQMC with PH framework, enhancing its generality and applicability through broader PH coverage and improved quality.

Yet, challenges remain.
To begin with, similar to ECPs, constructing high-quality PH depends on accurate reference data from experiments or high-level calculations, which can be costly to curate.
Besides, there is still room to improve the PH fitting scheme, which can benefit from careful hyperparameter tuning and the adoption of more general pseudopotential forms.
Furthermore, while the removal of core electrons improves convergence, simulating large and strongly correlated systems still demand careful initialization, robust optimization strategies, and expressive neural network ansatz.
Encouragingly, in addition to the NNQMC-motivated advances in PH, techniques developed for modern ECP construction are readily transferable to PH, and collaboration with the broader ECP community will accelerate the progress.

The potential of PH extends far beyond the systems investigated in this study. 
A key next step is the development of transferable Pseudo-Hamiltonians for a wider range of elements.
Furthermore, it is also straightforward to apply NNQMC with PH to real solids, by integrating PH into, for instance, DeepSolid model~\cite{li2022ab, qian2025deep, gerard2025transferable}.
This approach also hold promise for excited state calculation when combined with penalty method~\cite{ lu2023penalty, shepard2022double, entwistle2023electronic, li2024spin} or NES-VMC~\cite{pfau2024accurate}.
Additionally, our work can enhance the potential-energy-surface calculations using NNQMC~\cite{schatzle2025ab, scherbela2022solving, scherbela2024towards, gaoab, gaosampling, gao2023generalizing}.
Finally, by incorporating more efficient neural network ansatz, as recently proposed~\cite{FiRE}, NNQMC with PH offers highly accurate and scalable approaches for long-standing problems in physics and chemistry, such as the P-cluster~\cite{li2019electronic} and high-temperature superconductors \cite{zhou2021high}.

\section{Methods}
\subsection{Neural Network-based Quantum Monte Carlo}\label{NNQMC}
NNQMC is a class of highly accurate methods for solving the time-independent Schr\"{o}dinger equation~\cite{foulkes2001quantum}
\begin{equation}
\hat{H} \psi=E \psi,
\label{eq:sch_eq}
\end{equation}
where $\hat{H}$ is the Hamiltonian operator corresponding to the total energy of the system.
For realistic continuum systems, a Hamiltonian typically consists of kinetic and potential energy terms, as described in more detail in \cref{three_H}.
Regardless of the specific Hamiltonian form, NNQMC provides a unified framework: modeling the wavefunction with antisymmetric neural networks (to comply with Fermi-Dirac statistics) and solving the Schr\"{o}dinger equation via QMC methods such as VMC and DMC.

In VMC, the total energy serves as the training loss to obtain the first eigenstate (that is, the ground state) of \cref{eq:sch_eq}, based on the variational principle. 
Given a many-body wavefunction $\psi(\mathbf{x})$, where $\mathbf{x}=\mathrm{concat}\left(\mathbf{r}_1,\ldots,\mathbf{r}_n\right)$ represents the coordinates of electrons, the total energy is calculated as
\begin{equation}
    E\left[\psi\right] = \frac{\bra{\psi}\hat{H}\ket{\psi}}{\braket{\psi}} = \mathbb{E}_{\mathbf{x}\sim p(\mathbf{x})}\left[E_L\right],
    \label{eq:energy}
\end{equation}
where $p(\mathbf{x})=\frac{\left|\psi(\mathbf{x})\right|^2}{\int\left|\psi(\mathbf{x})\right|^2\mathrm{d}\mathbf{x}}$ is the square of normalized wavefunction, and $E_L(\mathbf{x})=\frac{\hat{H}\psi(\mathbf{x})}{\psi(\mathbf{x})}$ is the local energy.
Also, the corresponding gradient can be stochastically estimated as
\begin{equation}
    \nabla_\theta E\left[\psi\right] \propto \mathbb{E}_{\mathbf{x}\sim p(\mathbf{x})}\left[\left(E\left[\psi\right]-E_L\right)\nabla_\theta\log \left|\psi\right|\right],
\end{equation}
where $\theta$ represents parameters of the wavefunction to be optimized.

Under the variational principle, the accuracy of VMC is inherently limited by the expressiveness of the wavefunction ansatz.
DMC, however, as a projection-based method, overcomes this limitation.
It extracts the ground state $\psi_0$ by applying the imaginary time evolution operator $e^{-\tau\hat{H}}$ to an initial trial wavefunction $\psi$.
As $\tau\to\infty$, the projection $e^{-\tau\hat{H}}\psi$ exponentially supresses excited states, leaving $\psi_0$ as the dominant component.
This process is simulated using an ensemble of walkers that diffuse through configuration space (mimicking kinetic energy) and undergo population adjustments based on the potential energy landscape~\cite{carlo1990diffusion}.
For fermionic systems, the method relies on the fixed-node approximation to prevent sign problem~\cite{anderson1975random}.

\subsection{Different Hamiltonian Forms}\label{three_H}
Under Born-Oppenheimer approximation, the all-electron ab initio Hamiltonian in a molecular system is given by:
\begin{equation}
\begin{aligned}
\hat{H}=-\frac{1}{2} \sum_{i} \nabla_{i}^{2}-\sum_{I} \sum_{i} \frac{Z_{I}}{\left|\mathbf{r}_{i}-\mathbf{R}_{I}\right|} \\
+\sum_{i<j} \frac{1}{\left|\mathbf{r}_{i}-\mathbf{r}_{j}\right|}+\sum_{I<J} \frac{Z_{I} Z_{J}}{\left|\mathbf{R}_{I}-\mathbf{R}_{J}\right|},
\label{eq:H}
\end{aligned}
\end{equation}
where $i, j$ are subscripts for electrons, and $I, J$ for nuclei.
$Z_I$ denotes the charges, and $\mathbf{r}_i, \mathbf{R}_I$ denote the positions of electrons and nuclei respectively.

For heavy atoms, the intricate interplay of core electron dynamics leads to significant computational challenges.
However, predicting chemical properties usually hinges more critically on tackling valence electrons.
In such cases, an effective Hamiltonian form that eliminates core electrons can be used for simplification, and additional potential terms are added to mimic the interaction between core electrons and valence electrons, which reads as
\begin{align}
  \hat{H}^\mathrm{ECP} & = \hat{H}_\mathrm{valence}+\sum_I\sum_v V^\mathrm{ECP}_I(r_{vI}), \\
  V^{\mathrm{ECP}}(r) & =V_\mathrm{local}^\mathrm{ECP}(r)
  + \sum_{\ell=0}^{M-1} \sum_{m=-\ell}^{\ell} V_\ell(r) {\ket{\ell m}\bra{\ell m}}.
  \label{eq:semi}
\end{align}
Here, $\hat{H}_\mathrm{valence}$ retains the form of \cref{eq:H} but includes only valence electrons, and $r_{vI}=\left|\mathbf{r}_{v}-\mathbf{R}_{I}\right|$ represents the radial distance between valence electron $v$ and atom $I$.
The additional potential terms with respect to atom $I$ is denoted by $V_I^\mathrm{ECP}$, whose details are demonstrated in \cref{eq:semi}.
The potential terms $V_\mathrm{local}^\mathrm{ECP}$ and $V_\ell$ are usually expanded in Gaussian basis sets with parameters to be optimized, and $\ket{\ell m}$ represents the spherical harmonics.
Unfortunately, the nonlocal nature of the second term in \cref{eq:semi} adds significant computational overhead to the NNQMC calculation~\cite{li2022fermionic}. 

An alternative approach, known as Pseudo-Hamiltonian (PH), 
not only approximates the interactions between the ionic core and the valence electrons but also modifies the kinetic energy terms~\cite{bachelet1989novel} with a local effective term.
Under spherical symmetry around each atom, the general form of the PH is rigorously given by
\begin{align}  \hat{H}^\mathrm{PH}&=\hat{H}_\mathrm{valence}+\sum_I\sum_v h^\mathrm{PH}_I(r_{vI}), \label{eq: ph hamiltonian}\\
  h^\mathrm{PH}(r) &= \frac{1}{2}\hat{p}a\left(r\right)\hat{p} + V_\mathrm{local}^\mathrm{PH}(r) + V_{L^2}(r)\hat{L}^2,
  \label{eq:ph}
\end{align}
where $\hat{p}$ is the momentum operator and $\hat{L}$ is the angular momentum operator, and $a$, $V_\mathrm{local}^\mathrm{PH}$, $V_{L^2}$ are parameterized functions to be optimized, same as $V_\mathrm{local}^\mathrm{ECP}$ and $V_\ell$.
We refer to \cref{PH:construct} for the details of optimization.
It is worth noting that besides the kinetic energy, the local term $V_{L^2}(r)\hat{L}^2$ in \cref{eq:ph} contains a second-order differential operator, which could be computationally costly.
Fortunately, with a modified Forward Laplacian framework, this term can be calculated with negligible computational overhead.
See details in \cref{sec:fl} and Supplementary Note 2.

In this work, results labeled ``ECP" adopt ccECP~\cite{annaberdiyev2018new} for metal elements and sulfur, while those labeled ``PH" use the pseudo-Hamiltonian from Ref.~\cite{ichibha2023locality, bennett2022high} for metal elements and our own for sulfur.
The neural network we use for modeling the wavefunction is LapNet~\cite{li2024computational}, an efficient transformer-based NNQMC ansatz with high accuracy, with the training hyperparameters provided in Supplementary Note 1. 

\subsection{Pseudo-Hamiltonian Construction} \label{PH:construct}
In this section, we describe our PH construction procedure, using sulfur as an example.
When optimizing the pseudo-Hamiltonian, the electron's radial effective mass is fixed to be equal
to its actual mass by setting $a\left(r\right)=1$ following Ref.~\cite{ichibha2023locality, bennett2022high}.
Under this constraint, the Hamiltonian's kinetic energy terms reduce to the standard form, shifting focus on potential energy terms in construction.
Excluding the kinetic energy term, the pseudo-Hamiltonian can be expanded in terms of the spherical harmonics $\left|\ell m\right\rangle$ as follows:
\begin{equation}
\begin{aligned}
  &V^\mathrm{PH}(r) \\
  &= V_\mathrm{local}^\mathrm{PH}(r) + V_{L^2}(r)\hat {L}^2
   \\
  &= V_\mathrm{local}^\mathrm{PH}(r)+ V_{L^2}(r)
  \sum_{\ell=0}^{\infty} \sum_{m=-\ell}^{\ell} \ell (\ell + 1) {\left|\ell m\right>\left<\ell m\right|}.
  \label{eq:l2}
\end{aligned}
\end{equation}
By exploiting the similarity to ECP, the parameterized functions in PH are assigned such that the matrix elements match ECP's ones:
\begin{equation}
\begin{aligned}
\left<\ell m\left|V^\mathrm{PH}(r)\right|\ell m\right>
&=\left<\ell m\left|V\mathrm{^{ECP}}(r)\right|\ell m\right>,
\label{eq:matrix} \\
(\ell=0,1,\cdots,M-1;\;\;m&=-\ell, -\ell+1, \cdots, +\ell).
\end{aligned}
\end{equation}
Here, $M-1$ denotes the maximum angular momentum channel in the ECP's potential terms.
Specifically, the parameters were optimized as follows:
\begin{enumerate}
    \item Optimize the pseudo-Hamiltonian to reproduce the atomic properties and S$_2$ binding curve calculated with the ccECP using the Hartree--Fock (HF) method to minimize the cost function (see Supplementary Note 6).
    \item Based on the optimized pseudo-Hamiltonian, evaluate the binding curve of S$_2$ and FeS dimers , and compute the mean absolute error (MAE) with respect to the ccECP results using the CCSD(T) method.
    \item Update the weights of the cost function by Bayesian optimization to minimize the MAE values obtained in step~2.
    \item Repeat steps~1--3 for 50 iterations.
    \item Based on the best weights obtained, replace some of the terms in the cost function with an approximate evaluation at CCSD(T) accuracy, and re-optimize the pseudo-Hamiltonian. 
\end{enumerate}

The parameter functions of the PH were defined with reference to the BFD pseudopotential's 
~\cite{burkatzki2007energy,2008MB_MD}, following equation \ref{eq:matrix},
with the cutoff radius $r_c$ fixed at 2.195 Bohr (to match the Ne‑core ccECP value).
The pseudopotential consists of $s$, $p$, and $d$ channels, so $M-1=2$.
For more details in the optimization, see Supplementary Note 6.

\subsection{Pseudo-Hamiltonian calculation with Forward Laplacian}\label{sec:fl}
In this section, we demonstrate that the PH integration introduces almost no additional computational overhead to NNQMC under the Forward Laplacian framework.

\newcommand{\termone}[0]{\sum_{i,\alpha,\beta}A_{\alpha\beta}(\vr_i)\phat_{i, \alpha}\phat_{i, \beta}}

The Hamiltonian with PH used in this paper is given by \cref{eq: ph hamiltonian}. To better demonstrate the calculation process associated with Forward Laplacian, we rewrite the Hamiltonian according to the degree of $\bmphat_{i}$:
\begin{align}
        \hat{H}^{\mathrm{PH}}=& \termone + \sum_{i,\alpha}b_{\alpha}(\vr_i)\phat_{i,\alpha} \nonumber\\
        &\quad\quad + \Vall,\label{eq: hamiltonian pr form}
\end{align}
where $\alpha,\beta\in\{x,y,z\}$, $\phat_{i,\alpha}$ denotes the $\alpha$-axis momentum operator of the $i$-th electron. 
$\mA(\vr)$, ${\bf{b}}(\vr)$ and $\Vall$ are derived from \cref{eq: ph hamiltonian}, with explicit form provided in Supplementary Note 7. 
The first term in \cref{eq: hamiltonian pr form}, that is, $\termone$, necessitates the second-order operator calculation of neural networks:
\begin{equation}
\label{eq: second order operator}
    \sum_{i,\alpha,\beta}A_{\alpha\beta}(\vr_i)\partial_{r_{i\alpha}}\partial_{r_{i\beta}}\psi,
\end{equation}
which is rarely considered in existing NNQMC studies. 

To apply the Forward Laplacian framework~\cite{li2024computational}, we consider the following coordinate transformation:
\begin{equation}
\label{eq: coordinate transformation}
    \mathbf{v}_{i}=\boldsymbol{Q}_i^{-1}\mathbf{r}_{i},
\end{equation}
where $i=1,2,...,N$, 
given decomposition $\boldsymbol{A}(\mathbf{r}_i)=\boldsymbol{Q}_{i}^{\top}\boldsymbol{Q}_i$ with $\boldsymbol{Q}_i\in\mathbb{R}^{3\times 3}$ being invertible constant matrices. 
Based on the chain rule, we have:
\begin{equation}
    \sum_{i\alpha}\partial_{v_{i\alpha}}^2\psi = \sum_{i\alpha\beta}{A}_{\alpha\beta}(\mathbf{r}_i)\partial_{r_{i\alpha}}\partial_{r_{i\beta}}\psi.
    \label{eq: chain rule}
\end{equation}
Therefore, the second order operator in \cref{eq: second order operator} is transformed to the Laplacian with respect to $\mathbf{v}_i$, accelerated by the Forward Laplacian framework naturally. As shown in Ref.\cite{li2024dof}, this coordination transformation can be easily integrated with Forward Laplacian by modifying the input of the algorithm. Moreover, as the coordination transformation defined by \cref{eq: coordinate transformation} is an electron-wise transformation, the intrinsic sparse derivative structure in LapNet and other neural network-based ansatz remains the same, making the additional computational cost associated with psedo-Hamiltonian negligible.

\section*{Data Availability}
The dataset is available at Zenodo~\cite{Fu2026}. Source Data are provided with this paper.

\section*{Code Availability}
We open-source our implementation of NNQMC with PH approach in JaQMC repository on Github (https://github.com/bytedance/jaqmc), and the specific version code is archived at Zenodo (ref.~\cite{fu_2026_20323938}).

\section*{Acknowledgements}
We thank ByteDance Seed Group for inspiration and encouragement, and Hang Li for his guidance and support.
Some of the calculations reported in this paper were performed using the facilities of Center for Advanced Scientific Computing at JAIST.
Ji Chen is supported by the National Key R\&D Program of China (2021YFA1400500) and National Science Foundation of China (12334003).
Liwei Wang was supported by the National Science Foundation of China (NSFC92470123, NSFC62276005).
Tom Ichibha is supported by the JSPS KAKENHI Grant Number 
24K17618 and JSPS Overseas Research Fellowships.
Ryo Maezono is grateful for financial supports from MEXT-KAKENHI (22H051462, 24K01172, 24K07571A). 
Kenta Hongo is grateful for financial support from MEXT-KAKENHI, Japan (24K07571).
Ryunosuke Fujimaru appreciates support from JST BOOST, Japan (JPMJBS2425).

\section*{Author Contributions}
L.W., T.I., R.M., J.C. and W.R. conceived and supervised the project. W.F., R.L., X.L. and W.R. proposed and implemented the algorithms. W.F., Y.L., X.W. and W.R. performed and analyzed the simulations. R.F., K.H., T.I., R.M. constructed the local pseudopotentials. All the authors wrote the paper.

\section*{Competing Interests}
The authors declare no competing interests.

\appendix

\bibliography{ref}%
\end{document}

\title{ }

\author{ }

\date{\vspace{-5ex}}
\maketitle
\tableofcontents
\newpage

\section{Hyperparameters}
All calculations in the main text adopt default hyperparameters provided in Supplementary Table \ref{hp}, except for Iron-sulfur clusters training, whose details are presented in Supplementary Note \ref{Iron-S}.
\begin{table*}[!htbp]
\caption{\textbf{Default hyperparameters used in NNQMC calculations.} This table lists the network architecture, training, pretraining and KFAC settings used for the main-text calculations unless otherwise specified.}\label{hp}
\centering
\begin{tabular}{ccc}
  \toprule
   & Parameter & Value \\
  \midrule
  \multirow{5}*{LapNet} & Determinants & 16 \\
   & Network layers & 4 \\
   & Attention heads & 4 \\
   & Attention dimension & 64 \\
   & MLP hidden dimension & 256 \\
  \midrule
  \multirow{7}*{Training} & Optimizer & KFAC \\
   & Iterations & 2e5 \\
   & Learning rate at iteration t & $l_0 / \left(1+\frac{t}{t_0}\right)$ \\
   & Initial learning rate $l_0$ & 0.05 \\
   & Learning rate decay $t_0$ & 1e4 \\
   & Local energy clipping & 5.0 \\
   & MCMC decorrelation steps & 30 \\
  \midrule
  \multirow{3}*{Pretrain} & Iterations & 1e3 \\
   & Optimizer & LAMB \\
   & Learning rate & 1e-3 \\
  \midrule
  \multirow{4}*{KFAC} & Norm constraint & 1e-3 \\
  & Damping & 1e-3 \\
  & Momentum & 0 \\
  & Covariance moving average decay & 0.95 \\
  \bottomrule
\end{tabular}
\end{table*}

\section{Negligible Additional Computational Cost on PH}
Here we provide the training time cost of the three methods versus the number of electrons in Supplementary Figure~\ref{fig:time}, with data corresponding to Fig.~2a in the main text.
One sulfur atom contains 16 electrons in total, while only 6 valence electrons would be considered in PH/ECP calculations.
The PH and AE curves align closely, confirming that PH introduces negligible additional computational cost.
In contrast, ECP requires significantly more training time for the same number of electrons due to computationally intensive spherical harmonics integrations.

\begin{figure}[ht]
    \centering
    \includegraphics{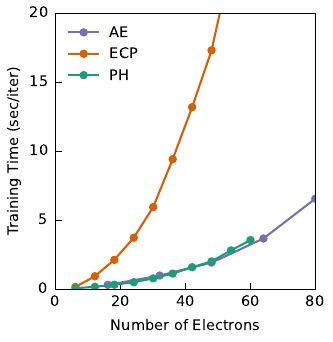}
    \caption{\textbf{Training time with respect to the electron number.} The plot compares the time cost of the AE, ECP and PH approaches for the systems in Fig.~2a of the main text, with the x-axis expressed as the number of explicitly considered electrons rather than the number of S atoms.}
    \label{fig:time}
\end{figure}

\section{Efficiency Comparison on Various Systems}
The NNQMC efficiency comparison among AE, ECP and PH Hamiltonians across various sulfur-containing, phosphorus-containing, chlorine-containing and metal-atom systems is presented in Supplementary Figures~\ref{fig:cluster_time},~\ref{fig:pcl_cluster_time} and~\ref{fig:metal_time}, respectively.
Across all systems, the qualitative trend is consistent: the computational cost of NNQMC-ECP is comparable to or even exceeds that of NNQMC-AE, while NNQMC-PH is substantially more efficient than both.
ECP being as expensive as AE may seem counterintuitive, as it reduces the number of explicit electrons.
However, it introduces expensive nonlocal projector evaluations, which become particularly costly when used with neural network wavefunctions, a point also emphasized in the main text.
The adoption of the Forward Laplacian technique~\cite{li2024computational} greatly reduces the cost of the kinetic energy evaluation, thereby shifting the computational bottleneck to these nonlocal ECP terms.

Notably, the performance gap between AE and ECP is more pronounced in metal-atom systems.
Conversely, for elements lighter than sulfur (e.g., Na, Mg, Al), where the core-to-valence electron ratio is larger, ECP is expected to regain its computational advantage over AE.
However, PH's superior efficiency relative to ECP is expected to persist across all such systems.
\begin{figure}[ht]
    \centering
    \includegraphics[width=125mm]{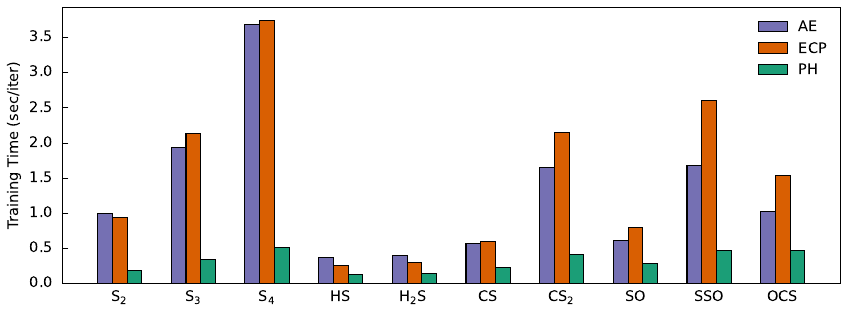}
    \caption{\textbf{NNQMC efficiency on sulfur-containing clusters.} The plot compares the computational cost of NNQMC calculations using AE, ECP and PH Hamiltonians for sulfur cluster systems.}
    \label{fig:cluster_time}
\end{figure}
\begin{figure}[ht]
    \centering
    \includegraphics[width=125mm]{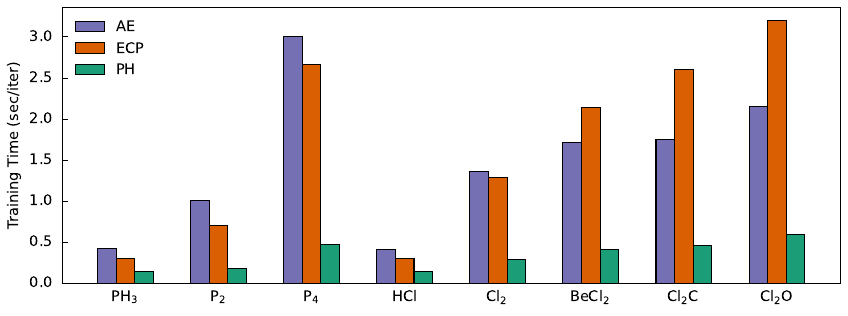}
    \caption{\textbf{NNQMC efficiency on phosphorus- and chlorine-containing clusters.} The plot compares the computational cost of NNQMC calculations using AE, ECP and PH Hamiltonians for clusters containing phosphorus or chlorine atoms.}
    \label{fig:pcl_cluster_time}
\end{figure}
\begin{figure}[ht]
    \centering
    \includegraphics[width=125mm]{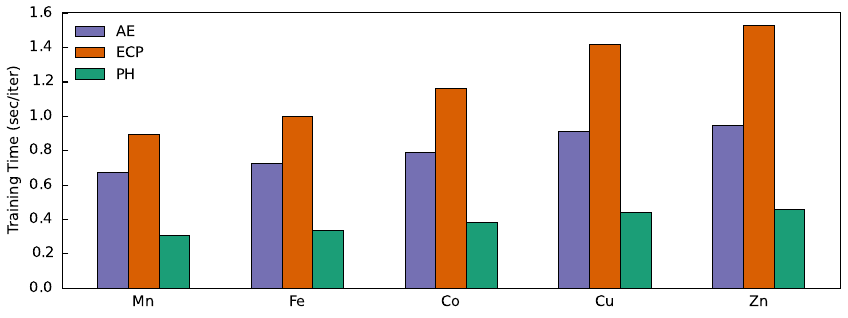}
    \caption{\textbf{NNQMC efficiency on metal-atom systems.} The plot compares the computational cost of NNQMC calculations using AE, ECP and PH Hamiltonians for the metal-atom systems studied in this work.}
    \label{fig:metal_time}
\end{figure}

\section{Atomization Energy Convergence Curves}
The full atomization energy convergence curves of sulfur, phosphorus and chlorine clusters, systems discussed in Fig.~3b and Fig.~4b in the main text, are shown in Supplementary Figure~\ref{fig:s_clusters}.
As evident from the curves, NNQMC calculations with AE Hamiltonian after fivefold longer training have nearly approached convergence.
Nevertheless, the accuracy of NNQMC-AE remains inferior to that of both NNQMC-ECP and NNQMC-PH across all systems.

\begin{figure}[ht]
    \centering
    \includegraphics[width=130mm]{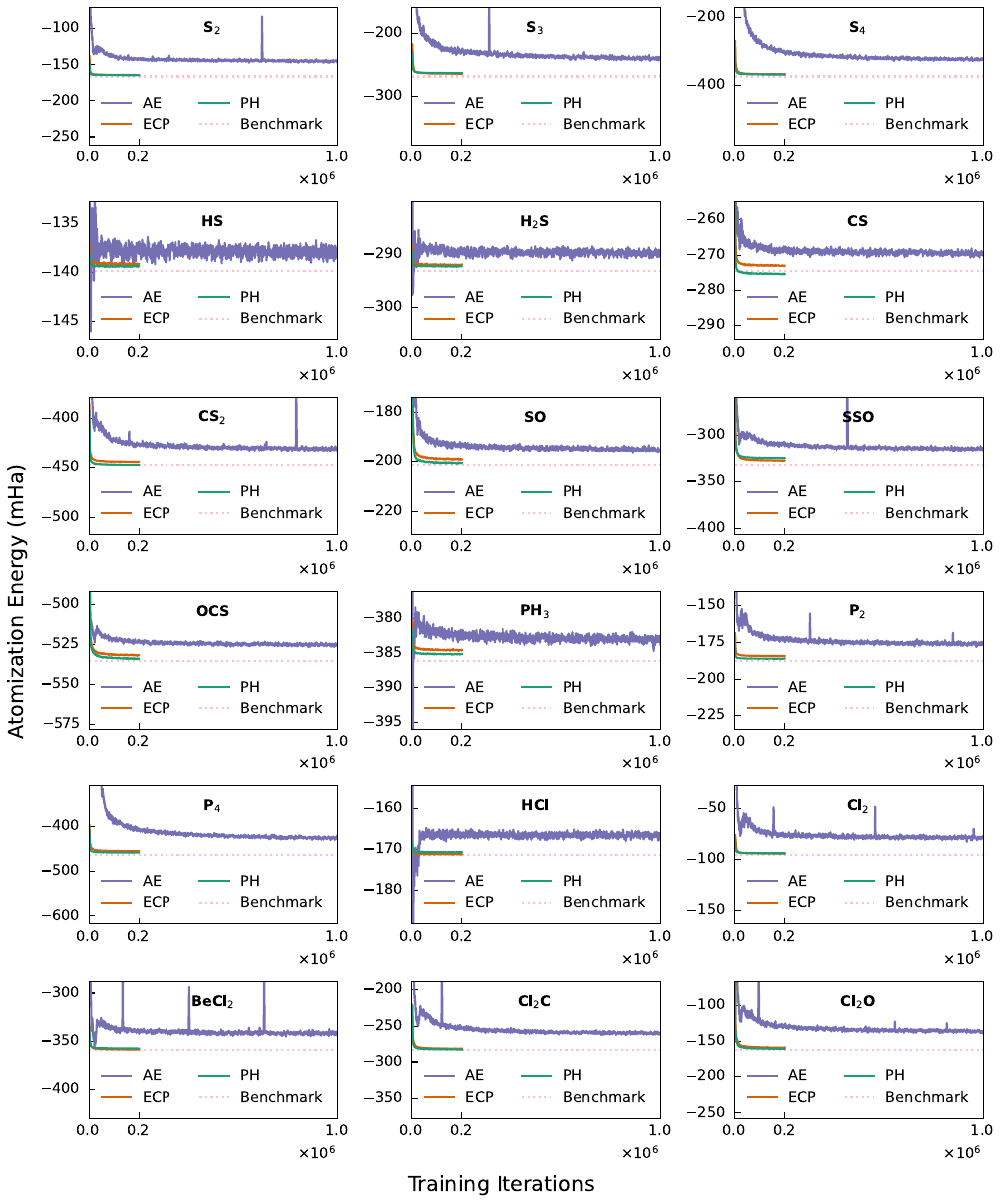}
    \caption{\textbf{Atomization energy convergence curves for sulfur, phosphorus and chlorine clusters.} The plot shows the full NNQMC training curves, where atomization energies are obtained by subtracting isolated-atom energies from total cluster energies to allow comparison across Hamiltonians. All curves are smoothed with a 2000-step rolling window.}
    \label{fig:s_clusters}
\end{figure}

\section{Strong Correlation Effects of Sulfur Dimer}
In addition to the potential energy curves (PECs) of triplet states presented in Fig. 3c of the main text, we further investigated the PECs of singlet states, as detailed in Supplementary Figure~\ref{fig:sg}.
To benchmark our results, we performed NEVPT2 calculations—a high-accuracy correlated method.
Notably, the PECs obtained using the ECP/PH approach closely align with the NEVPT2 reference results, whereas the AE method exhibits significant deviations beyond 2.7 Å.
This discrepancy mirrors the trends observed in the main text and underscores the limitations of AE in strongly correlated regimes for sulfur dimer.

\begin{figure}[ht]
    \centering
    \includegraphics{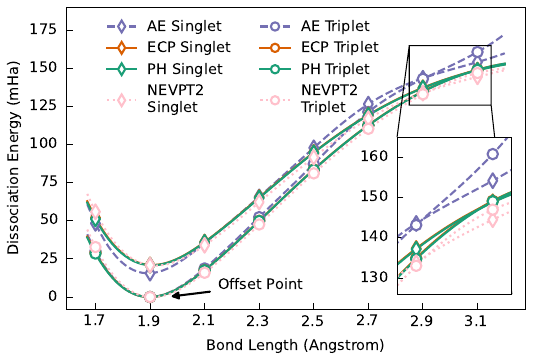}
    \caption{\textbf{Potential energy curves of singlet and triplet sulfur dimer states.} The plot compares the singlet and triplet energy curves, using the triplet energies at 1.9~\AA~as reference points for energy offsets. All curves are corresponding spline curves for guiding the eye. The inset highlights the crossover region.}
    \label{fig:sg}
\end{figure}

\section{Construction of the Pseudo-Hamiltonian}
The transformation from the semi-local pseudopotential to PH is as follows \cite{2007MB_MD,2018MCB_LM}:
\begin{align}
\left<\ell m\left|V^\mathrm{PH}\right|\ell m\right>
&=\left<\ell m\left|V\mathrm{^{ECP}}\right|\ell m\right>,
\label{eq:matrix} \\
(\ell=0,1,2;\;\;m&=-\ell, -\ell+1, \cdots, +\ell). \nonumber
\end{align}

However, for the solutions of the parameter functions
$V_\mathrm{local}^\mathrm{PH}$ and $V_{L^2}$,
the parameter functions of the semi-local pseudopotential 
must satisfy the following linear dependency condition:
\begin{equation}
  \label{eq:linear}
  2 V_0(r) - 3 V_1(r) = 0.
\end{equation}
Moreover, the following bounding condition, which is a unique constraint for the pseudo-Hamiltonian, must also be satisfied~\cite{1989GBB_MGBC}:
\begin{equation}
  \label{eq:bounding}
  1+2r^2 V_{L^2}(r)>0.
\end{equation}
This condition is imposed to keep the electron's rotational effective mass non-negative.
If it is not satisfied, the kinetic energy decreases as the electron approaches the core,
causing the wave function to collapse into a delta-function distribution around the nucleus while lowering the energy.

Ultimately, the optimization of the pseudo-Hamiltonian reduces to optimizing a semi-local pseudopotential that satisfies the constraints \eqref{eq:linear} and \eqref{eq:bounding}.
With respect to the linear dependency condition in Eq.~\eqref{eq:linear}, since sulfur has the 3$p$ orbitals as its valence shell,
it is undesirable to alter the parameter function for the $p$ channel. Therefore, we defined $V_0(r)$ as a functional of $V_1(r)$ by setting
\[
V_0[V_1(r)] = \frac{3}{2}V_1(r),
\]
so that Eq.~\eqref{eq:linear} is always satisfied.
Regarding the bounding condition in Eq.~\eqref{eq:bounding},
we incorporate a corresponding constraint term into the cost function, as will be described later,
and, following the previous studies \cite{2023TI_FAR,2022MCB_JTK,2020JTK_FAR},
we add a correction term consisting of a linear combination of Gaussian functions to $V_{L^2}(r)$,
which provides the necessary degrees of freedom to satisfy the condition.
We denote the modified function including the correction term by ${V'}_{L^2}(r)$. The correction term for ${V'}_{L^2}(r)$ is composed of a linear combination of three Gaussian functions
with fixed exponents 3.0, 5.0 and 7.0 Bohr$^{-2}$, with their coefficients treated as optimizable parameters. 

In the training process, the cost function is given by:
\begin{equation}
   \mathcal{O}\equiv\omega_1 O_{\mathrm{eigenenergy}}+\omega_2 O_{\mathrm{norm}}+\omega_3 O_{\mathrm{excitation}}^{\mathrm{HF}}
   +\omega_4 O_{\mathrm{S_2 binding}}^{\mathrm{HF}}+\lambda O_{\mathrm{bounding}}
    \label{eq:cost}
\end{equation}
where 

\begin{align}
  &\mathcal{O}_{\mathrm{eigenenergy}} = \sum_{i}\left[\frac{\varepsilon_{i}^{\mathrm{PH}}-\varepsilon_{i}^{\mathrm{ccECP}}}{\varepsilon_{i}^{\mathrm{ccECP}}}\right]^2,
  \label{eq:o-eigenenergy} \\
  &\mathcal{O}_{\mathrm{norm}} = \sum_{i}\left[\left\langle\phi_{i}^{\mathrm{PH}}\middle|\phi_{i}^{\mathrm{PH}}\right\rangle_{r>r_{c}}-\left\langle\phi_{i}^{\mathrm{ccECP}}\middle|\phi_{i}^{\mathrm{ccECP}}\right\rangle_{r>r_{c}}\right]^2,
  \label{eq.o-norm} \\
  &\mathcal{O}_{\mathrm{excitation}}^{\mathrm{HF}} = \sum_{j\in(3s,3p)}\left[\frac{\Delta E_{j}^{\mathrm{HF},\mathrm{PH}}-\Delta E_{j}^{\mathrm{HF},\mathrm{ccECP}}}{\Delta E_{j}^{\mathrm{HF},\mathrm{ccECP}}}\right]^2,
  \label{eq.o-excitation} \\
  &\mathcal{O}_{\mathrm{S_2 binding}}^{\mathrm{HF}} = \sum_{d}\left[\Delta E_{d}^{\mathrm{HF},\mathrm{PH}}-\Delta E_{d}^{\mathrm{HF},\mathrm{ccECP}}\right]^2.
  \label{eq.o-binding} \\
  &\mathcal{O}_{\mathrm{bounding}} = \int dr\left[\mathcal{R}_s^{\delta b',\delta b'_s}\left(2r^2{V'}_{L^2}(r)\right)-2r^2{V'}_{L^2}(r)\right],
  \label{eq:o-bounding}
\end{align}

where $\Delta E_{j}$ represents the electron excitation energies and $\Delta E_{d}$ denotes the binding energies for different distances. The terms $\mathcal{O}_{\mathrm{eigenenergy}}$ and $\mathcal{O}_{\mathrm{norm}}$ correspond to the reproduction of the valence orbital eigenenergies and the norms outside the cutoff radius, respectively, for the ground state configuration $3s^2\,3p^4$. The term $\mathcal{O}_{\mathrm{excitation}}^{\mathrm{HF}}$ becomes lower as the electron excitation energies are more accurately reproduced. For the excited states, high-spin states ranging from charge -1 to +5, as well as any lower spin states (if any), were considered. The term $O_{\mathrm{S_2 binding}}^{\mathrm{HF}}$ decreases as the binding curve of the reference S$_2$ dimer is better reproduced. The term $\mathcal{O}_{\mathrm{bounding}}$ is a penalty term to enforce the bounding condition of Eq.~\eqref{eq:bounding}. The function $\mathcal{R}_s^{\delta b',\delta b'_s}$ is a smooth and shifted ramp function \cite{2020JTK_FAR} that transforms the given function $2r^2{V'}_{L^2}(r)$ so that it does not fall below -1 for all $r$ (thus ensuring the bounding condition is met). If $2r^2{V'}_{L^2}(r)$ already satisfies the bounding condition, no modification is applied. Consequently, the integral in Eq.~\eqref{eq:o-bounding} increases as $2r^2{V'}_{L^2}(r)$ increasingly violates the bounding condition. The parameters $\delta b'$ and $\delta b'_s$ adjust the shape of $\mathcal{R}_s^{\delta b',\delta b'_s}$; following our previous studies \cite{2022MCB_JTK,2023TI_FAR}, we set $\delta b'=0.25$~Ha and $\delta b'_s =5\times 10^{-3}$~Ha.

In the cost function for optimization step 4, the terms $\mathcal{O}_{\mathrm{excitation}}^{\mathrm{HF}}$ and $O_{\mathrm{S_2 binding}}^{\mathrm{HF}}$ in Eq.~\eqref{eq:cost} are replaced by the following expressions, thereby incorporating the correlation energy at an approximate CCSD(T) level into each term:

\begin{align}
    &\mathcal{O}_{\mathrm{excitation}}^{\mathrm{CCSD(T)}}
    = \sum_{j \in (3s, 3p)} \left[
    \frac{ \Delta E_{j}^{\mathrm{HF,PH}} - \Delta E_{j}^{\mathrm{CC}} }
    { \Delta E_{j}^{\mathrm{CC}} }
    \right]^2, \label{eq:ex_ccopt} \\
    &\mathcal{O}_{\mathrm{S_2\,binding}}^{\mathrm{CCSD\left(T\right)}}=
	\sum_{d}\left[\Delta E_{d}^{\mathrm{HF},\mathrm{PH}}-\Delta E_{d}^{\mathrm{CC}}\right]^2, \label{eq.binding_ccopt} \\
    &\Delta E^{\mathrm{CC}} \equiv \Delta E^{\mathrm{CCSD(T),ccECP}} - \Delta E^{\mathrm{corr,PH}}, \label{eq:shift} \\
    & \Delta E^{\mathrm{corr,PH}} \simeq \Delta E^{\mathrm{CCSD(T),HFOPT}} - \Delta E^{\mathrm{HF,HFOPT}}
\end{align}

Here, $\Delta E^{\mathrm{CC}}$ is defined as the difference between the reference CCSD(T) energy, $\Delta E^{\mathrm{CCSD(T),ccECP}}$, and the correction term $\Delta E^{\mathrm{corr,PH}}$, which converts $\Delta E^{\mathrm{HF,PH}}$ into $\Delta E^{\mathrm{CCSD(T),PH}}$. In this work, $\Delta E^{\mathrm{corr,PH}}$ is approximated by the value evaluated for the PH obtained in optimization step~3 (denoted as HFOPT).
This approach avoids the need to perform computationally expensive CCSD(T) calculations
for each parameter update of the pseudo-Hamiltonian.
Please refer to the previous studies\cite{2022MCB_JTK,2023TI_FAR} for further details.

For the evaluation of the valence orbital eigenenergies, norms, and electron excitation energies during optimizing the pseudo-Hamiltonian,
we employed a numerical atomic Hartree--Fock code \cite{1997SK_CWCa,1997SK_CWCb}.
$\Delta E^{\mathrm{CC}}$ in Eq.~\eqref{eq:shift} and the calculation 
of the binding curves were carried out using the Molpro code \cite{2011HJW_MS,2020HJW_MS}.
In Molpro, fully uncontracted aug-cc-pwCVnZ basis sets (with n = T, Q, and 5) were used\cite{2002KAP_TMDJ}.
For the binding curve calculations, a quadruple-zeta basis set was employed,
which has been confirmed to yield sufficient accuracy \cite{2023TI_FAR}.
Meanwhile, the electron excitation energies were evaluated at the complete basis set (CBS) limit
by extrapolating using basis sets with $n$ = T, Q, and 5, according to the following formulas \cite{2018AA_LM}:
\begin{align}
  &E_{n}^{\mathrm{HF}} = E_{\mathrm{CBS}}^{\mathrm{HF}} + a\,\exp\left[-bn\right], \\
  &E_{n}^{\mathrm{corr}} = E_{\mathrm{CBS}}^{\mathrm{corr}} + \frac{c}{\left(n+3/8\right)^3} + \frac{d}{\left(n+3/8\right)^5}.
\end{align}

Here, $E_{n}^{\mathrm{HF}}$ and $E_{n}^{\mathrm{corr}}$ are the HF energy and the correlation energy, respectively, obtained with an $n$-zeta basis set.
The correlation energy is defined as the difference between the CCSD(T) energy and the HF energy.

The minimization of the cost function was performed using the Nelder--Mead method \cite{1962WS_FRH,1965JAN_RM} implemented in SciPy \cite{2020PV_S1C}.
For Bayesian optimization, we employed the Hyperopt library \cite{2015JB_DDC} with the Tree-structured Parzen Estimator algorithm \cite{2011JB_BK}.
The conversion of the pseudopotential format and the computations related to the bounding condition were carried out
using the Nexus workflow management code \cite{2016KRO}.

\section{Details of Integrating Forward Laplacian with PH}
In this section, we show how to derive the main text Eq. (11) from the main text Eq. (6). We begin with expanding main text Eq. (6) into:
\begin{align}  \hat{H}^\mathrm{PH}&=\hat{H}_\mathrm{valence}+\sum_I\sum_v h^\mathrm{PH}_I(r_{vI})\\
&=\sum_{v,\alpha}\frac{1}{2}\hat{p}_{v,\alpha}\hat{p}_{v,\alpha}-\sum_{v,I}\frac{Z_I}{r_{vI}}+\sum_{v< v'}\frac{1}{|\br_v-\br_{v'}|} + \sum_{I<J}\frac{Z_IZ_J}{|{\bf{R}}_I-{\bf{R}}_J|}+ \sum_{v,I}V_\mathrm{local,I}^\mathrm{PH}(r_{vI}) + V_{L^2,I}(r_{vI})\hat L^2_{vI},
\end{align}
where $\hat L_{vI}={\bf r}_{vI}\times \hat{p}_v$ denotes the angular momentum operator of the $i$-th electron with respect to the $I$-th nuclei, ${\bf r}_{vI}={\bf r}_{v}-{\bf R}_{I}$ denotes the relative position between the $i$-th electron and $I$-th nuclei. As $\hat L_{vI}^2$ does not contains zeroth-order term with respect to $\hat p_v$, the $V_{\mathrm{all}}$ in main text Eq. (11) is defined by:
\begin{equation}
    V_{\mathrm{all}}(\br_1,\br_2,...,\br_N)=-\sum_{v,I}\frac{Z_I}{r_{vI}}+\sum_{v< v'}\frac{1}{|\br_v-\br_{v'}|} + \sum_{I<J}\frac{Z_IZ_J}{|{\bf{R}}_I-{\bf{R}}_J|}+ \sum_{v,I}V_\mathrm{local,I}^\mathrm{PH}(r_{vI}),
\end{equation}
which includes all the potential energy terms in $\hat{H}_{\mathrm{valence}}$ and all the local potential terms of the PH. Then, $\hat{H}^\mathrm{PH}$ can be simplified to:
\begin{align}
\hat{H}^\mathrm{PH}&=\sum_{v,\alpha}\frac{1}{2}\hat{p}_{v,\alpha}\hat{p}_{v,\alpha}+\sum_{v,I}V_{L^2,I}(r_{vI})\hat L^2_{vI} + V_{\mathrm{all}}(\br_1,\br_2,...\br_N)\\
&=\sum_{v,\alpha}\frac{1}{2}\hat{p}_{v,\alpha}\hat{p}_{v,\alpha}+\sum_{v,I}V_{L^2,I}(r_{vI})\left(\sum_{\alpha\beta}(r_{vI}^2\delta_{\alpha\beta}- r_{vI,\alpha}r_{vI,\beta})\hat{p}_{v,\alpha}\hat{p}_{v,\beta}+2\mathbbm{i} \sum_{\alpha}r_{vI,\alpha}\hat{p}_{i,\alpha}\right) + V_{\mathrm{all}}(\br_1,\br_2,...\br_N)\\
&=\sum_{v,\alpha,\beta}\left(\frac{1}{2}\delta_{\alpha\beta}+\sum_{I}V_{L^2,I}(r_{vI})(r_{vI}^2\delta_{\alpha\beta}- r_{vI,\alpha}r_{vI,\beta})\right)\hat{p}_{v,\alpha}\hat{p}_{v,\beta}+\sum_{v,\alpha}2\mathbbm{i}\sum_{I}V_{L^2,I}(r_{vI})r_{vI,\alpha}\hat{p}_{i,\alpha}+V_{\mathrm{all}}(\br_1,\br_2,...\br_N).
\end{align}
Here $\mathbbm{i}$ is the imaginary unit. Thus, we can define $A_{\alpha\beta}(\br)$ and $b_{\alpha}(\br)$ according to 
\begin{align}
    A_{\alpha\beta}(\br)&=\frac{1}{2}\delta_{\alpha\beta} +\sum_{I}V_{L^2,I}\left(|\br-{\bf R}_I|\right)\left(|\br-{\bf R}_I|^2\delta_{\alpha\beta}-(r_{\alpha}-{ R}_{I,\alpha})(r_{\beta}-{ R}_{I,\beta})\right),\\
    b_{\alpha}(\br)&=2\mathbbm{i}\sum_{I}V_{L^2,I}\left(|\br-{\bf R}_I|\right)(r_{\alpha}-{ R}_{I,\alpha}).
\end{align}

\section{Iron-sulfur cluster training details}\label{Iron-S}
In iron-sulfur cluster systems,  NNQMC often faces multiple local minima during training LS state due to complex spin structures. To efficiently obtain broken-symmetry states, we recommend using broken-symmetry UHF (BS-UHF) as the pretraining target. This approach produces smoother training curves and achieves lower energy compared to standard RHF/UHF pretraining. For the $[\text{Fe}_2\text{S}_2(\text{SCH}_3)_4]^{2-}$, We train the LS state through the BS-UHF pretraining target and obtain the meaningful magnetic coupling parameters $J$.

We conducted training for 200,000 steps on the BS and HS of $[\text{Fe}_2\text{S}_2(\text{SCH}_3)_4]^{2-}$, and for $[\text{Fe}_4 \text{S}_4 (\text{SCH}_3)_4]$, we performed 500,000 steps of VMC training followed by DMC training. The detailed energies can be found in Supplementary Table \ref{e_fe}.

\begin{table*}[!htbp]
\centering
\caption{\textbf{Energies of iron--sulfur clusters.} This table reports VMC and, where available, DMC total energies for the broken-symmetry (BS) and high-spin (HS) states of $[\text{Fe}_2\text{S}_2(\text{SCH}_3)_4]^{2-}$ and $[\text{Fe}_4 \text{S}_4 (\text{SCH}_3)_4]$.}\label{e_fe}
\begin{tabular}{ccc}

  \toprule
    & VMC & DMC \\
  \midrule
  $[\text{Fe}_2\text{S}_2(\text{SCH}_3)_4]^{2-} \ (\text{BS})$   & -467.8407(3)   &-   \\
  \midrule
  $[\text{Fe}_2\text{S}_2(\text{SCH}_3)_4]^{2-} \ (\text{HS})$   & -467.8147(3)   & -   \\
  \midrule
  $[\text{Fe}_4 \text{S}_4 (\text{SCH}_3)_4] \ \ \ \ \ (\text{HS})$   & -734.8450(2)   & -734.8918(7)   \\
  \bottomrule
\end{tabular}
\end{table*}

\section{Iron-sulfur cluster Reduced Density Matrix analysis}
To analyze the Spin structure, we should calculate the one-body reduce density matrix(1-RDM) and two-body reduce density matrix(2-RDM). 
1-RDM $\rho^{\sigma}_{ij}$  and 2-RDM $\rho^{\sigma_1 \sigma_2}_{ijkl}$ can be given by:
\begin{align}
   \rho^{\sigma}_{ij} &= \left< \hat{\rho}^{\sigma}_{ij} \right > = \left< \hat{c}^{\dagger}_{\sigma, i} \hat{c}_{\sigma, j}\right> \\
   \rho^{\sigma_1 \sigma_2}_{ijkl} &= \left< \hat{\rho}^{\sigma_1 \sigma_2}_{ijkl} \right > = \left< \hat{c}^{\dagger}_{\sigma_1, i}\hat{c}^{\dagger}_{\sigma_2, k} \hat{c}_{\sigma_2, l} \hat{c}_{\sigma_1, j}
   \right>
\end{align}
where $c^{\dagger}_{\sigma i}(c_{\sigma i})$ represent the creation and annihilation operators on the $i$th orbital with $\sigma$ spin. 

In the VMC framework, 1-RDM and 2-RDM can be calculated by mapping wavefunction to orbitals. Detailed calculation details are described in the paper\cite{wheeler2023pyqmc}.

The spin operator can be written:
\begin{align}
    \hat{S}_{i}^{x} &= \frac{1}{2} (\hat{c}^{\dagger}_{i \uparrow}\hat{c}_{i \downarrow} 
                    +\hat{c}^{\dagger}_{i\downarrow}\hat{c}_{i\uparrow} ) \\
    \hat{S}_{i}^{y} &= \frac{i}{2} (\hat{c}^{\dagger}_{i \downarrow}\hat{c}_{i \uparrow} 
                    -\hat{c}^{\dagger}_{i\uparrow}\hat{c}_{i\downarrow} ) \\
    \hat{S}_{i}^{z} &= \frac{1}{2}(\hat{c}^{\dagger}_{i\uparrow}\hat{c}_{i\uparrow} -
    \hat{c}^{\dagger}_{i\downarrow}\hat{c}_{i\downarrow}) = \frac{1}{2} (\rho^{\uparrow}_{ii} - 
    \rho^{\downarrow}_{ii}) \\
    \hat{S}_{i}\hat{S}_{j} &= \hat{S}^{x}_{i}\hat{S}^{x}_{j} + \hat{S}^{y}_{i}\hat{S}^{y}_{j} + \hat{S}^{z}_{i}\hat{S}^{z}_{j}
\end{align}

Next, we expand $S_{i}S_{j}$. For clarity, we divide it into two cases: $i = j $ and $i \neq j$.  If $i = j$:
\begin{equation}
\begin{aligned}
     \hat{S}^{2}_{i} &= \frac{3}{4}(\hat{c}^{\dagger}_{i\uparrow}\hat{c}_{i\uparrow}+\hat{c}^{\dagger}_{i\downarrow}
     \hat{c}_{i\downarrow}-2\hat{c}^{\dagger}_{i\uparrow}
     \hat{c}_{i\uparrow}\hat{c}^{\dagger}_{i\downarrow}\hat{c}_{i\downarrow}) \\
     &= \frac{3}{4} (\hat{\rho}^{\uparrow}_{ii} + \hat{\rho}^{\downarrow}_{ii}-2 
     \hat{\rho}^{\uparrow \downarrow}_{iiii}) 
\end{aligned}
\end{equation}

If $i \neq j$:
\begin{equation}
\begin{aligned}
    \hat{S}_{i}\hat{S}_{j} &= -\frac{1}{2}(\hat{c}^{\dagger}_{i\uparrow}\hat{c}_{j\uparrow}\hat{c}^{\dagger}_{j\downarrow}
    \hat{c}_{i\downarrow}+
    \hat{c}^{\dagger}_{i\downarrow}\hat{c}_{j\downarrow}\hat{c}^{\dagger}_{j\uparrow}\hat{c}_{i\uparrow})+\frac{1}{4}(\hat{c}^{\dagger}_{i\uparrow}\hat{c}_{i\uparrow}\hat{c}^{\dagger}_{j\uparrow}\hat{c}_{j\uparrow} + 
    \hat{c}^{\dagger}_{i\downarrow}\hat{c}_{i\downarrow}\hat{c}^{\dagger}_{j\downarrow}\hat{c}_{j\downarrow}
    -2\hat{c}^{\dagger}_{i\uparrow}\hat{c}_{i\uparrow}\hat{c}^{\dagger}_{j\downarrow}\hat{c}_{j\downarrow}) \\
    & = -\frac{1}{2}(\hat{\rho}^{\uparrow\downarrow}_{ijji} + \hat{\rho}^{\downarrow\uparrow}_{ijji}) 
    + \frac{1}{4}(\hat{\rho}^{\uparrow\uparrow}_{iijj} + \hat{\rho}^{\downarrow\downarrow}_{iijj}-2\hat{\rho}^{\uparrow\downarrow}_{iijj})
\end{aligned}
\end{equation}
As is well known, the magnetic properties of iron-sulfur clusters arise from the orbitals of the iron atoms. Therefore, when calculating the spin properties, we only consider the $3d$, $4s$ and $4d$ of Fe orbitals by L\"{o}wdin orthogonalization. 

The local spin $\hat{S}^{z}_A$, local spin square $\hat{S}^{2}_A$ and spin correlation $\hat{S}_{A}\hat{S}_{B}$ can be defined 
\begin{equation}
    \begin{aligned}
    &\hat{S}^{z}_A = \sum_{i \in A}\frac{1}{2}\hat{S}^{z}_i \\
    &\hat{S}^{2}_A = \sum_{i, j\in A, i\neq j} \hat{S}_{i}\hat{S}_{j} + \sum_{i \in A}\hat{S}^{2}_{i} \\
    &\hat{S}_A\hat{S}_{B} = \sum_{i\in A, j \in B}  \hat{S}_{i}\hat{S}_{j}
    \end{aligned}
\end{equation}

In the case of the HS state of $[\text{Fe}_4 \text{S}_4 (\text{SCH}_3)_4]$, the Heisenberg model predicts $S^{2}_A=8.75$ and $\left<S_{A}S_{B}\right>=6.25$, which differs from the results of the NNQMC calculation shown in the Fig. 4e of main text. This discrepancy can be attributed to the transfer of spin and charge from iron to sulfur. This phenomenon has also been discussed in the referenced paper~\cite{sharma2014low} 

\bibliography{si-bibliography}